\def\year{2022}\relax
\documentclass[letterpaper]{article} 
\usepackage{aaai22}  
\usepackage{times}  
\usepackage{helvet}  
\usepackage{courier}  
\usepackage[hyphens]{url}  
\usepackage{graphicx} 
\usepackage{natbib}  
\usepackage{caption} 
\DeclareCaptionStyle{ruled}{labelfont=normalfont,labelsep=colon,strut=off} 
\usepackage{algorithm}
\usepackage{algorithmic}

\usepackage{newfloat}
\usepackage{listings}
\floatstyle{ruled}
\newfloat{listing}{tb}{lst}{}
\floatname{listing}{Listing}
\title{Socially Fair Mitigation of Misinformation on Social Networks via\\ Constraint Stochastic Optimization}
\author {
    Ahmed Abouzeid,\textsuperscript{\rm 1}
    Ole-Christoffer Granmo,\textsuperscript{\rm 1} 
    Christian Webersik,\textsuperscript{\rm 2}
    Morten Goodwin\textsuperscript{\rm 1}
}
\affiliations{
\textsuperscript{\rm 1}Centre for Artificial Intelligence Research (CAIR), University of Agder, Norway\\
\textsuperscript{\rm 2}Centre for Integrated Emergency Management (CIEM), University of Agder, Norway\\
\{ahmed.abouzeid, ole.granmo, christian.webersik, morten.goodwin\}@uia.no
}

\usepackage{bibentry}
\begin{document}

\maketitle

\begin{abstract}
Recent social networks' misinformation mitigation approaches tend to investigate how to reduce misinformation by considering a whole-network statistical scale. However, unbalanced misinformation exposures among individuals urge to study fair allocation of mitigation resources. Moreover, the network has random dynamics which change over time. Therefore, we introduce a stochastic and non-stationary knapsack problem, and we apply its resolution to mitigate misinformation in social network campaigns. We further propose a generic misinformation mitigation algorithm that is robust to different social networks' misinformation statistics, allowing a promising impact in real-world scenarios. A novel loss function ensures fair mitigation among users. We achieve fairness by intelligently allocating a mitigation incentivization budget to the knapsack, and optimizing the loss function. To this end, a team of Learning Automata (LA) drives the budget allocation. Each LA is associated with a user and learns to minimize its exposure to misinformation by performing a non-stationary and stochastic walk over its state space. Our results show how our LA-based method is robust and outperforms similar misinformation mitigation methods in how the mitigation is fairly influencing the network users. 
\end{abstract}

\section{Introduction}
From a computation perspective, there are many approaches to combat the dissemination of misinformation\footnote{The term misinformation is sometimes used to refer to all forms of fake news/content. However, in some literature, misinformation is defined as the unintentional spread of false content while disinformation is the on-purpose spread. In this paper, we refer to all forms of false content as misinformation.}. Recently, \cite{de2020approaches} illustrated some of the main techniques for classifying misinformation content and how these approaches can be applied in different scenarios. However, classification methods tend to be offline and limited to particular social network features to be learned, such as linguistics and the local political context \cite{lazer2018science}. Furthermore, such classification models have a potential for \textit{False Positive} matches, which may violate human rights conventions by misjudging and questioning individuals credibility and controlling free speech \cite{ozdan2021right}. On the other hand, recent work proposed intervention-based resolutions as an online approach to mitigate the circulation of misinformation on social media platforms. Such an approach is considered more convenient since it facilitates better collaboration between humans and technology by providing learned misinformation mitigation strategies instead of black-box classification models. For example, \cite{farajtabar2017fake} proposed a reinforcement learning-based optimization method which provides a strategy to decrease the difference between misinformation and true content exposures in Twitter, given that such misinformation exposure was dominating the network. The purpose was to mitigate the effect of misinformation on network users by incentivizing the latter to spread true information. A similar method was developed to facilitate decentralized and faster computation, as proposed by \cite{abouzeid2021learning}.

The latter approach introduces a light-weight decentralized computation that reduces the optimization sample space and utilizes Learning Automata (LA) that learn from reinforcement feedback \cite{narendra1974learning}. However, the method was evaluated according to the decrease in difference between the dominating misinformation and the incentivized true content, averaging over the whole network. The problem with such an approach is that there would be real-world scenarios where some individuals need mitigation efforts more than others, while a sub-network individuals would be already protected from high misinformation exposures. Therefore, we believe a more socially fair intervention and allocation of mitigation resources should be introduced under the framework of \cite{abouzeid2021learning}.    

This paper proposes a robust LA-based decentralized mitigation method that addresses a wide range of possible unbalanced exposures to either misinformation or true content, seeking robustness on a variety of a social network's statistics. Our contribution is threefold:

\begin{itemize}
\item We propose a novel learning scheme for an LA learning in a stochastic and non-stationary environment. The randomness comes from an information diffusion model based on point processes \cite{laub2015hawkes}, while the non-stationarity comes from the temporal changes that occur over the whole network when an individual user responds to incentivization. This non-stationarity is particularly intricate due to the hidden dependencies in the information diffusion model. The LA task is to construct a network of individual automata on top of the social network. Each individual automaton is associated with a single user and performs a constraint Knapsack optimization via a random walk \cite{pearson1905problem} over the automaton state space. 

\item We propose a novel loss function to ensure that true content incentivization budget is fairly assigned according to individual users exposure needs. To this end, the problem is defined as a stochastic and non-stationary multi-agent Knapsack \cite{nicosia2009multi} optimization problem.

\item We introduce two evaluation metrics (Achieved Mitigation and Achieved Fairness) to measure the efficiency and robustness of the proposed misinformation mitigation algorithm on different social network's statistics. And we evaluate how our proposed technique is more socially fair compared to the proposed approach in \cite{abouzeid2021learning}. We conduct our empirical experiments on both synthetic and real-world social networks. Software source code and data are available at (\url{https://github.com/Ahmed-Abouzeid/MMSS}).
\end{itemize}

\section{Preliminaries}

\subsubsection{Information Diffusion Modelling.}In order to apply intervention-based resolutions to misinformation mitigation, an information diffusion model is required to simulate the social network which to intervene with. The simulation is considered because intervention with the actual social media platforms is not feasible. We simulate the process of information diffusion by employing a Multivariate Hawkes process (MHP) as practiced by \cite{goindani2020social}, \cite{abouzeid2021learning}, and \cite{farajtabar2017fake}. An MHP is a multivariate stochastic process \cite{chen2016thinning} which models the occurrence of temporal or spatio-temporal asynchronous events by capturing the mutual-excitation (dependencies) between these events. To model the social network dynamics, each user is represented by two Hawkes processes (HP), one for misinformation dissemination behavior, and the other for true content. The associated user HPs generate estimated random counts for both information types, given some behaviour observation in the past (e.g., estimating number of re/tweeted events given historical dependency). These counts indicate the intensity of the process at a specific time realization. Hence, An HP can be defined with its conditional intensity function $\lambda$. The intensity function has two main components: base intensity $\mu$, and exponential decay function $g$ over an adjacency matrix  $A$. The formal explanation of the conditional intensity function is given by:

\begin{equation}
    \label{eq1}
    \lambda_{i}(t_r|H^{t_r}):= \mu_{i} + \sum_{t_{s<t_r}} g(t_r-t_s).
\end{equation}

Where $\mu$ is the base intensity that models some external motivation to propagate some content (independent from inferred relationships in data). On the other hand, $g$ is some kernel function over the observed history $H^{t_r}$ associated with user $i$ from the discrete time realization $t_{s}$ prior to time $t_r$. $g$ is concerned with the history of some influence matrix $A_{i.}$, where $A_{ij}$ = 1 if there is an influence indicating that user i influences user j, and $A_{ij}$ = 0 if not. We used an exponential decay kernel function $g=A_{i.} e^{-wt}$ as practiced by \cite{farajtabar2017fake}, where $w$ is the decay factor which represents the rate for how the influence is reduced over time. For all users, the base intensity vector $\mu$, and the influence matrix $A$ can be estimated using maximum likelihood as proposed in \cite{ozaki1979maximum}. To simulate all users behaviours for each content type, an MHP is created, given that different intensity rates are generated at different discrete time realizations. Hence, at each realization, each user behaviour is simulated as an estimated number of events (misinformation or true content) to be generated. We set the interval window between realizations to two hours. The HP simulation algorithm adopted in this study follows the modified thinning algorithm introduced by \cite{ogata1981lewis}. See Appendix A.1 for a detailed explanation of the simulation evaluation metric.

\subsubsection{Mitigated Diffusion.}The core idea behind misinformation mitigation is by introducing the true information to the network through incentivization. Therefore, users associated true content HPs are modified. Hence, let $x_i$ be the incentivization amount decided for user $i$, and the modified HP for mitigation purposes can be redefined by:

\begin{equation}
    \label{eq2}
    \lambda_{i}(t_r|H^{t_r}):= x_i + \mu_{i} + \sum_{t_{s}<t_r} g(t_r-t_{s}).
\end{equation}

\section{Related Work}
\subsubsection{Misinformation Impact.} According to \cite{bradshaw2017troops}, at least $50\%$ of the world's countries suffer from organized political manipulation campaigns over social media. Other examples of misinformation can be observed during the Ebola outbreak in West Africa, which was believed to be three times more worse than the previous Ebola outbreaks \cite{jin2014misinformation}. Therefore, research on the role of online media and border-free passing through messages became an emerging topic of interest in scientific communities. Furthermore, investigation on such a topic is more complicated and requires different perspectives of analysis. For example, recent studies \cite{rampersad2020fake} argued that the influence of social media on accepting political misinformation may differ depending on age, culture or gender. Such social studies actively investigated the social impact of misinformation propagation on different social media platforms such as Reddit, Facebook, and Twitter. Novel views on the problem emerged recently. For instance, recent investigations reported that deliberation contexts promoted in social media overcome false information about health \cite{pulido2020new}. An example of such deliberation can be viewed as a counterfactual campaigns to spread true health information against the spread of misinformation as practiced for the COVID-19 case on Twitter by \cite{abouzeid2021learning}. 

\subsubsection{Misinformation Detection.}The spread of fake news on social media has been initially considered as the intentional dissemination of false content in news articles \cite{allcott2017social}. Progressively, others gave attention to the broader range of the problem \cite{sharma2019combating,shu2017fake}. Moreover, rumor detection \cite{zhang2015automatic}, malicious accounts classification \cite{zannettou2019let,shao2018spread}, and the causal aspects of misinformation \cite{abouzeid2019causality} have been discussed. However, the majority of these methods are highly depending on linguistic or local features which cause a lack of generality in the final resolution. To the best of our knowledge, it is hard to solve the problem in real-time or without data selection-bias concerns \cite{ousidhoum2020comparative}. \cite{wasike2013social} expressed similar moral concerns since fake news detection resolutions are judgemental by nature. Therefore, the need for safer online strategies that would lead to more generic and authentic resolutions is critically desirable.
 
\subsubsection{Knapsack Optimization.}The utilization of Learning Automaton (LA) with Knapsack optimization problems is widely approached in the literature. For instance, \cite{granmo2007learning} worked on optimizing the allocation of polling resources for web page monitoring when the monitoring capacity is restricted. In web page monitoring systems, the system may involve $n$ web pages that are updated on different time intervals. Hence, to avoid involving all web pages including the ones with no updates, the system must determine the most important web pages only, without exceeding the monitoring capacity. The work utilized a team of learning automata, where each automaton is involved with a particular web page and learns its importance to a Knapsack total value. Similarly, \cite{yazidi2018solving} dealt with a Stochastic Non-linear Fractional Equality Knapsack (NFEK) problem which is a fundamental resource allocation problem based on incomplete and noisy information. In the latter work, they proposed an optimal resolution to the resource allocation problem using a continuous LA without mapping the Knapsack materials onto a binary hierarchy. In such work, the proposed LA had a Reward-Inaction (R-I) learning scheme which only updates the LA actions (transitions) probabilities when rewarded. \cite{ulker2017migrating} worked on another combinatorial optimization problem for Knapsack with a proposed Migrating Birds Optimization (MBO) algorithm to solve a 0-1  knapsack problem \cite{freville2004multidimensional}.

\subsubsection{Hawkes Processes.} The utilization of Hawkes processes-based intervention strategies was effectively presented on minimizing-risk problems. For example, \cite{gupta2018discrete} worked on the problem of invasive species spreading to new areas which threatens the stability of ecosystems and causes major economic losses. The latter study proposed a novel approach to minimize the spread of an invasive species given a limited intervention budget, where the spread of species was modelled by a Hawkes process and the minimization task was considered a constraint Knapsack optimization problem.

\section{Methodology}
\subsubsection{Learning Automata Network.}
\begin{figure}[t]
\centering
\includegraphics[width=0.55\columnwidth, height=280pt]{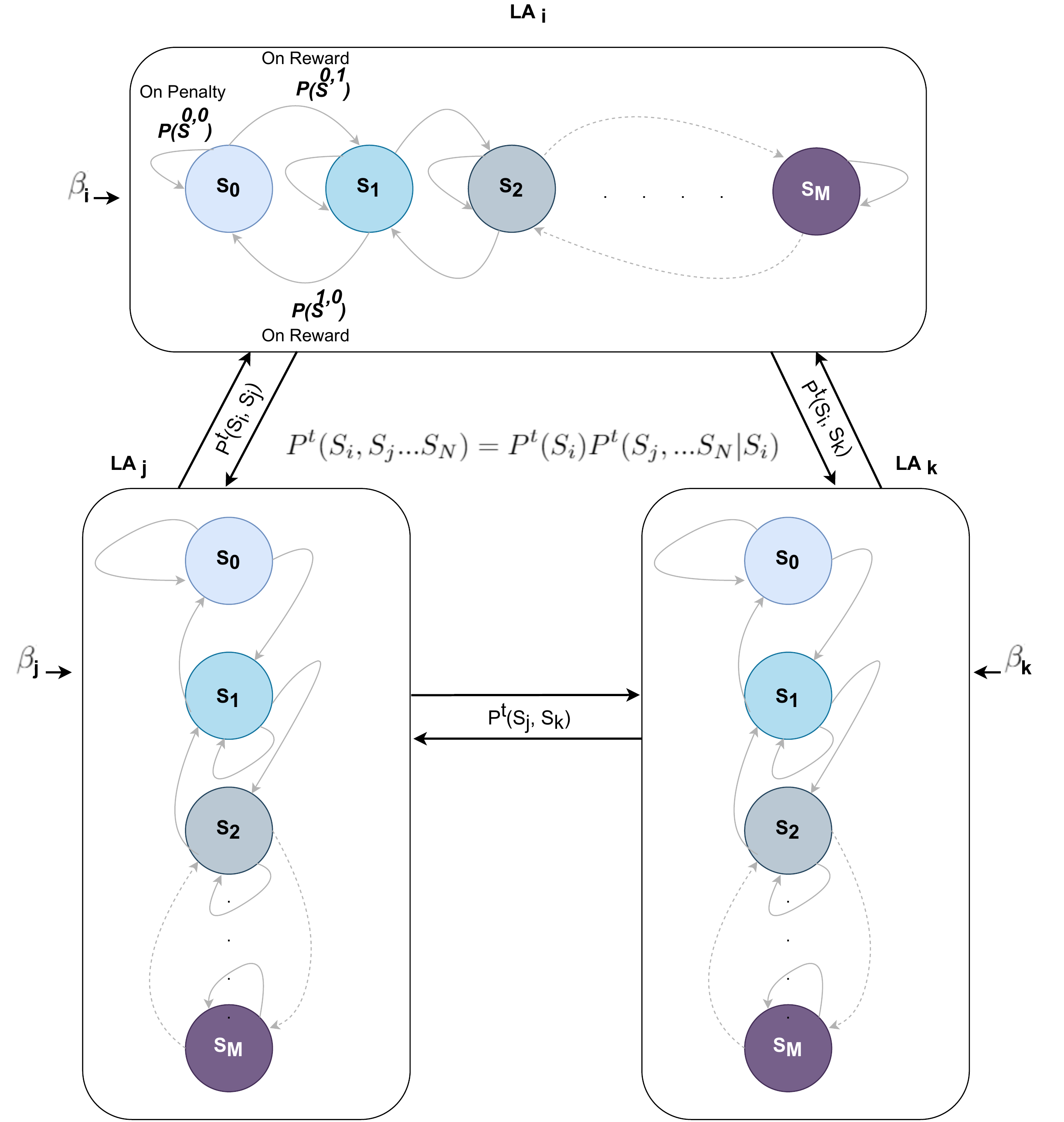} 
\caption{The proposed LA network and the underlying multivariate Markov chain architecture for three automata.}
\label{fig1}
\end{figure}

A Learning Automaton (LA) is a stochastic model suitable for learning in random environments \cite{narendra1974learning}. The LA learns by interacting with the random environment, and updates its actions or state transitions according to the stochastic signal from the environment. Depending on the automaton design and architecture, the task is to find either an optimum/sub-optimum action or state. The LA seeks convergence to such state or action, eventually. The advantage of utilizing an LA-based optimization is due to its decentralized and easy implementation. An LA defined by its stochastic state transitions can be formally defined as a Markov Process \cite{ames1989markov}. Therefore, to reach equilibrium over all LA, we build a network of LA, each performs a random walk over a finite and discrete state space, where the individual optimum or sub-optimum states will be the recommended incentivization values for a misinformation mitigation campaign. The individual random walks together form as a multidimensional joint random walk \cite{marquioni2019multidimensional} modelled by a multivariate Markov chain \cite{gotzamani2018introducing}. Figure \ref{fig1} demonstrates the proposed LA network and the underlying multivariate Markov chain (e.g., three automata with $M$ states, each.), where the joint state transitions and their probabilities are derived by the individual automata state transitions which are dictated by a reward signal $\beta$.

\subsubsection{Learning State Transition.} An individual $LA_i$ has a state space with memory depth $M$, where $M>0$. If $LA_i$ is in a state $S_{i}^k$ where $0 < k < M$, then, it has a three possible state transitions: $S_{i}^{k,k-1}, S_{i}^{k,k}, S_{i}^{k,k+1}$ indicating going to left, staying at same state, and moving to the right, respectively. In order to reach an optimum or sub-optimum state $S_{i}^*$, $LA_i$ needs to learn the probabilities of its state transitions until it converges. Consequently, the optimum or sub-optimum $S_{i}^*$ value will be the recommended incentivization value $x_{i}^*$ to modify the information diffusion model with (See Equation 2). $LA_i$ could have only two possible state transitions: $S_{i}^{k,k}, S_{i}^{k,k+1}$ or $S_{i}^{k,k}, S_{i}^{k,k-1}$, when $k=0$ or $k=M$, respectively. At each interaction step $t$, the probability of $LA_i$ being in a next state depends on its present state and the transition direction $a_{i}^{t}$. With a uniform initial state transitions probabilities, $LA_i$ determines the next state $S_{i}^{t+1}$ and updates its state transition probability distribution vector $\pi_{i}$ according to the below:

\begin{equation}
        \delta : S_{i}^{t}, a_{i}^{t}, \beta_{i}^{t} \to  S_{i}^{t+1}, \pi_{i}^{t+1}.
    \label{eq3}
\end{equation}

Where $\pi_{i}$ states probabilities are updated with regard to their rewarded visits frequency, and $a_{i}^{t}$ represents the applied state transition $a_{i}^{t} = S_{i}^{k,j}$, where $k, j$ are neighbor state indices and $k = j$ if it was a recurrent state transition. Based on $a_{i}^{t}$ and the environment stochastic reward $\beta_{i}^{t}$, $LA_i$ conducts a random step move over its state space. For instance, if $a_{i}^{t_0} = S_{i}^{k,k+1}$, the state transition function $\delta$ commits the transition $S_{i}^{k} \to S_{i}^{k+1}$ only if  $\beta_{i}^{t_0} = 0$, and rolls it back if $\beta_{i}^{t_0} = 1$. Consequently, $\pi_{i}^{t_1} = [0^{k,k}, 1^{k,k+1}, ...]^{t_1}$ or $\pi_{i}^{t_1} = [1^{k,k}, 0^{k,k+1}, ...]^{t_1}$, respectively. We denote $v_{i}^t$ and $w_{i}^t$ as how many times a transition was rewarded ($\beta_i = 0$) and performed for $LA_i$ up to interaction step $t$, respectively. Hence, For the state indices $k, j$, when $k = j+1$, state transition probabilities are updated as the below:
\begin{equation}
    P^{t+1}(S_{i}^{k,j}) = \frac{v_{i}^{t}(S_{i}^{k,j})}{w_{i}^{t}(S_{i}^{k,j})}, 
    \label{eq4}
\end{equation}

\begin{equation}
    P^{t+1}(S_{i}^{j,k}) = \frac{1-P^{t+1}(S_{i}^{k,j})}{2},
    \label{eq5}
\end{equation}

\begin{equation}
    P^{t+1}(S_{i}^{k,k}) = \frac{1-P^{t+1}(S_{i}^{k,j})}{2}, 
    \label{eq6}
\end{equation}

\begin{equation}
\mbox{where } P^{t+1}(S_{i}^{k,j}) + P^{t+1}(S_{i}^{j,k}) + P^{t+1}(S_{i}^{k,k}) = 1.
    \label{eq7}
\end{equation}

Since each LA performs a random walk over its state space through a stochastic state transition, then the optimization problem is solved by the multidimensional joint random walk over the automata network. Furthermore, the individual state transitions are dependent to each others due to the shared knapsack capacity and the inter-connected influence in their environment rewards. Therefore, the probability of a particular automata network state is calculated as the joint probability of the individual automata current states. Hence the joint probability can be calculated as the below, where N is the network size:

\begin{equation}
    P^{t}(S_{i}, S_{j} ... S_{N}) = P^{t}(S_{i}) P^{t}(S_{j}, ... S_{N}|S_{i}).
    \label{eq8}
\end{equation}

\subsubsection{LA Environment.}
To learn incentivization values for the social network's users, all users' associated LA interact with a Knapsack which evaluates how valuable the current LA state (incentive) for the mitigation campaign. The Knapsack evaluation is individual to each user behaviour on the network. Users behaviours are modeled through a multivariate Hawkes process (MHP). Hence, the LA environment has the following main properties.

\begin{itemize}
\item \textbf{Stochastic: } which is due to the randomness of each HP itself, which generates random counts for each  user events (e.g., re/tweets).
\item \textbf{Non-stationary: } which occurs because of the dependencies between users HP generated events. For instance, when both users $i, j$ have an explicit or implicit dependency, a particular incentivization value $x_i=0$ might not be optimum for user $i$ but could be optimum when the incentivization value $x_j>0$. Since the latter could cause user $i$ to be fairly exposed to true content without the need to increase for $x_i$ (incentivize user $i$).
\end{itemize}

To reinforce the learning of targeted state values. each individual $LA_i$ will receive a reward signal $\beta_i$ from its Knapsack environment where $\beta_i \in \{1, 0\}$, indicating a penalty, or reward Knapsack signal, respectively. The final committed state transition for an $LA_i$ is driven by the reward signal $\beta_i$. For instance, if $LA_i$ randomly walks towards the right and received a reward, it commits the transition and updates its current state. However, if $LA_i$ receives a penalty, it rolls back the transition and stays at its recent current state before that transition. The state update mechanism also works if $LA_i$ randomly walks to the left direction. These random walks probabilities in both directions are learned according to Equations \ref{eq4}, \ref{eq5}. On the other hand, recurrent state transitions probabilities are updated according to Equation \ref{eq6} until converging to a state where the probabilities of performing random walks in both directions became almost 0. The detailed information about how the reward signal $\beta_i$ is calculated for each direction of an $LA_i$ random walk:

\begin{equation}
    \label{eq9}
    \mbox{($\rightarrow$) }
    \beta_i(\textit{$m_i$}, \Phi) := 
        \left\{\begin{array}{lr}
        1, & \mbox{if } \textit{ $m_i$ } > 0 \lor \Phi = 1\\
        0, & \mbox{otherwise}\\
        \end{array}\right\},
\end{equation}

\begin{equation}
    \label{eq10}
        \mbox{($\leftarrow$) }
\beta_i(\textit{$m_i$}, \Phi) := 
        \left\{\begin{array}{lr}
        1, & \mbox{if } \textit{ $m_i$ } > 0 \\
        0, & \mbox{otherwise}\\
        \end{array}\right\},
\end{equation}
\begin{equation}
    \label{eq11}
    \mbox{subject }\mbox{to } \textit{ $m_i$ } = \frac{\Delta \mathcal{F}(x_i)}{\Delta x_i}, \mbox{ where } \Delta x_i > 0.
\end{equation}

\begin{figure}[t]
    \centering
   \includegraphics[width=0.55\columnwidth, height=200pt]{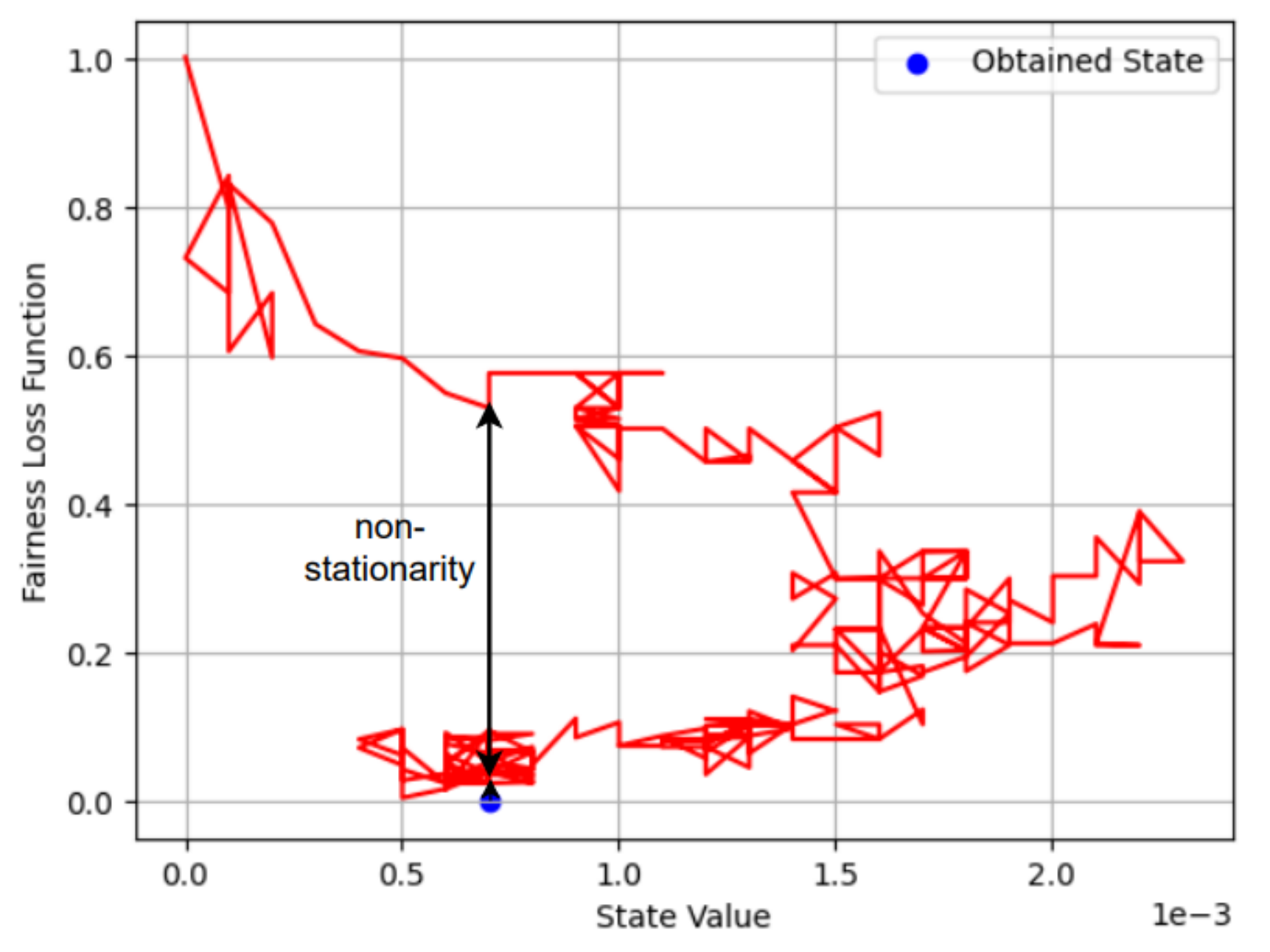}
   \caption{Finding global minima example for an individual LA random walk over a stochastic and non-stationary HP-based Knapsack response.}
   \label{fig2}
\end{figure}

Where \textit{ $m_i$ } is the slope of a fairness loss function $\mathcal{F}$ for the associated user $i$ and $\Phi$ indicates either the Knapsack is currently full ($\Phi = 1$) or not ($\Phi = 0$). The Knapsack initial capacity starts with $0$ and increased or decreased according to each individual LA state transition, while the current Knapsack capacity is shared across the LA network. Given that $x_i = S_i: i  \leq M$, since the mitigation incentive $x_i$ over time is represented by the current LA state where such an LA has M states. The above definition of the environment reward for the proposed random walk state transitions ensures converging to optimum or sub-optimum mitigation incentive values. Figure \ref{fig2} shows an example of our proposed LA state transitions mechanism where the optimization environment is non-stationary and stochastic. However, the LA managed to find a sub-optimal state value.

\subsubsection{Fairness Loss Function.}To achieve fair mitigation, we need to consider each individual user exposures to both misinformation and true content. Each user exposure associated with a content type is calculated as how much impact that content has on the user. Therefore, the ratio between true and misinformation impact for each user is considered. Hence, a more skewed initial distribution of these ratios will acquire a fair mitigation strategy to assign the incentivization budget according to user needs, without wasting the budget on users with already high exposures to true content. During the intervention, a ratio $R_i < 1$ means that user $i$ is more exposed to misinformation. Alternatively, a ratio $R_i > 1$ indicates that user $i$ incentivization is not necessary since the latter has already high level of true content exposures. The exposure values used in $R_i$ were calculated as proposed by \cite{abouzeid2021learning}, see Appendix A.2 for more details. Below, we define our proposed fair misinformation mitigation loss function:

\begin{equation}
    \label{eq12}
    \mbox{min } \mathcal{F}(X) := \sum_{i}^{N} \mathcal{F}(x_i), \mbox{ where }\mathcal{F}(x_i) := \sum_{j=0}^{n} (1 - R_{j}^{x_i})^2,
\end{equation}

\begin{equation}
    \label{eq13}
    \mbox{subject }\mbox{to } \sum^{N}_{i=1}x_{i}, \mbox{ where }  x_{i} \in [0, C].
\end{equation}

Where $N$ represents the number of network users and $n$ is the number of adjacent users connected to user $i$, where user $i$ is considered adjacent to itself. Therefore, $j$ is the index represents $i$ and all its adjacent over the summation. $R_{j}^{x_i}$ represents the updated ratio between true content and misinformation after applying the incentivization value $x_i$ to the true content HP diffusion model associated with user $i$. As noticed in Equation \ref{eq12}, we square the subtraction $1 - R_{j}^{x_i}$ to maintain positive values in the interval $[0, \infty)$, while the task is to minimize the loss function as much closer to 0 as possible (See Figure \ref{fig2}). It is important to highlight that the total loss is calculated through the achieved individual loss of each user during the allocation of incentives (e.g., associated LA and its current state value). That means the total loss ensures optimum or sub-optimum assigned incentivization values over $X$, where $X$ can be viewed as the set of all automata current states. Eventually, the consumption of all incentivization values (LA states) must not exceed the bound $C$, which represents the Knapsack capacity.  

\subsubsection{Misinformation Mitigation.}To obtain the optimum or sub-optimum learned states vectors of $N$ automata, we initialize each individual $LA_i$ with an initial state transition probability vector $\pi_{i}^{t_0}$, and the initial ratio $R^{x_i=0}$ where no incentivization values yet to be added to the associated estimated base intensity $\mu_{i}^{t_0}$ of the relevant HP. Eventually, the initial fairness loss function $\mathcal{F}_{i}^{t_0}(x_i=0)$ is calculated while the Knapsack is initially empty $c^{t_0}=0$. The mitigation algorithm then iterates over the whole LA network until it converges to all optimum or sub-optimum state probability vectors. Then, converged states values are suggested as incentivization values for the underlying associated users on the network. The details of the misinformation mitigation procedure is shown in Algorithm 1.
\begin{algorithm}[tb]
\textbf{Input: } $\mu_{i}^{t_0}, \pi_{i}^{t_0}, R^{x_i=0}, \mathcal{F}_{i}^{t_0}(x_i=0), \forall i: u_i \in U, c^{t_0}$, and $N$ where $|U|=N$. \\
\textbf{Output}: $S_{i}^{*}, \forall i : u_i \in U$, where $|S^*| = N$.\\
\begin{algorithmic}[1]
\STATE Let $t=1$.
 \WHILE{$\neg(\pi_{all}^t \leftarrow \pi_{all}^*)$}{
 \FOR{$i \leftarrow 1 $ to $N$}{
  \IF{$\pi_{i}^t \neq \pi_{i}^*$}
      \STATE $a_{i}^{t} \leftarrow max[P(S_{i}^{k,j}), P(S_{i}^{k,k}), P(S_{i}^{j,k})]^t$.\\
       \STATE $S_{i}^{t} \leftarrow a_{i}^{t}$. \\
       \STATE$x_i \leftarrow S_{i}^{t}$. \\
       \STATE$\Delta x_i \leftarrow abs(S_{i}^{t} - S_{i}^{t-1})$.\\
       \STATE  $\sum_{j=0}^{n}(R_{j}^{x_i}) \leftarrow \lambda(x_i)$.\\
       \STATE$\mathcal{F}^t(x_i) \leftarrow \sum_{j=0}^{n}(1-R_{j}^{x_i})^2$. \\
       \STATE$\Delta {F}(x_i) \leftarrow abs({F}(x_i)^{t} - {F}(x_i)^{t-1})$. \\
       \STATE $m_i = \frac{\Delta {F}(x_i) }{\Delta x_i}$.  \\
       \STATE$\beta_{i}^{t} \leftarrow \beta_{i}^{t}(m_i, \Phi)$. \\
       \STATE$S_{i}^{t+1}, \pi_{i}^{t+1} \leftarrow \delta(S_{i}^{t}, a_{i}^{t}, \beta_{i}^{t}) $.\\
   
   \ELSE
       \STATE $continue$.\\
  \ENDIF
  }
 \ENDFOR \\
  $t \leftarrow t+1$.
 }
\ENDWHILE
 \label{alg1}
\caption{Fair misinformation mitigation.}
\STATE \textbf{return} $S^*$.
\end{algorithmic}
\end{algorithm}

\section{Experimental Setup}
 In our experiments we design six synthetic social networks $\{syn1, syn2, syn3, ..., syn6\}$. Each with a unique statistical misinformation exposure distribution among users. The six networks represent the possible real-world scenarios where some user groups might be highly exposed to misinformation more than other groups on the social network. Moreover, some individuals in these groups might be also highly exposed to misinformation more than others from the same group. Allowing for these possible scenarios in our experiments should stress the evaluation of robustness for a fair misinformation mitigation resolution. We design our synthetic networks by randomly generate variant true information and misinformation event counts on both user and network levels. Then, we set different bounds on these synthetic exposures to maintain a variety of statistics for each network. Eventually, we run our resolution on a real-world social network used in \cite{abouzeid2021learning} as another benchmark. The real-world network is a COVID-19 social network and annotated for ordinary and false re/tweets from Twitter on the $28^{th}$ of March, 2020. The collected re/tweets focused on discussions about COVID-19. The criteria for the misinformation annotation was if any propagated content urged the public for using false drugs \cite{tesfaye2020we} without any official statements from the health authorities at that time. Within each of our experiments, we consider different mitigation incentivization budget for the Knapsack capacity to evaluate for different levels of constraints. Due to the randomness of experiments, we run each for multiple times and take the average as an estimate of the final outcome. Table \ref{tbl1} shows the configuration of our experiments, where all networks have 200 users. For the selection of hyper-parameters values in all experiments, see Appendix A.3.

 \begin{table}[t]
  \caption{Configuration details of fair misinformation mitigation experiments on the proposed social networks.}
  \centering
  
  \begin{tabular}{ccc}
      \hline
    \textbf{Network} & \textbf{Knapsack size}& \textbf{Overall Misinformation}\\
    \hline
        Syn1 & 0.06 & 17.00\% \\
        Syn2 & 0.06 & 58.00\%  \\
        Syn3 & 0.06 & 88.50\% \\
        COVID-19 & 0.06 & 89.50\%\\
        Syn4 & 0.18 & 11.75\%  \\
        Syn5 & 0.18 & 47.25\% \\
        Syn6 & 0.18 & 86.50\% \\
        COVID-19 & 0.18 & 89.50\%\\

        \hline
  \end{tabular}
  \label{tbl1}
\end{table}

\section{Evaluation}

\subsubsection{Uniform-baseline.}To highlight the need for a fair misinformation mitigation method, we make an analogy with a uniform allocation of the incentivization budget. For instance, if all or almost network users are equally exposed to misinformation than true content, a uniform distribution of incentivization budget is theoretically an optimum fair mitigation strategy. We refer to the latter as \textbf{Case-0}. However, the more the two content types were unbalanced on the network, the more challenging for a budget uniform distribution to achieve the desired mitigation results. For example if only $20\%$ of network users were exposed to misinformation, a uniform incentivization becomes a waste for $80\%$ of the budget, which might cause no mitigation at all since $20\%$ of the budget becomes insufficient to maintain $R=1$ for the targeted users. We refer to the latter as \textbf{Case-1}. Another form of skewness is when the majority of users are exposed to misinformation but a subset of them are significantly more exposed to misinformation than others, in such scenario, the uniform method will suffer as well, since these subset of users will need more incentivization than others. We refer to the latter as \textbf{Case-2}. It is important to highlight that the purpose of the HP information diffusion model is to predict future behaviours. Therefore, the initial distribution of misinformation exposures before any future intervention is unknown, and a robust incentivization is mandatory to overcome all the potential misinformation percentages.
\subsubsection{AVG-LA-baseline.} We further investigate how our LA-based resolution performs against current existed LA-based methods \cite{abouzeid2021learning}. We refer to the latter as \textbf{AVG-LA}, while we refer to our proposed method as \textbf{Fair-LA}.
\subsubsection{Mitigation Efficiency.} 
\begin{figure*}[h!]
\centering 
\includegraphics[height=7cm, width=8.6cm]{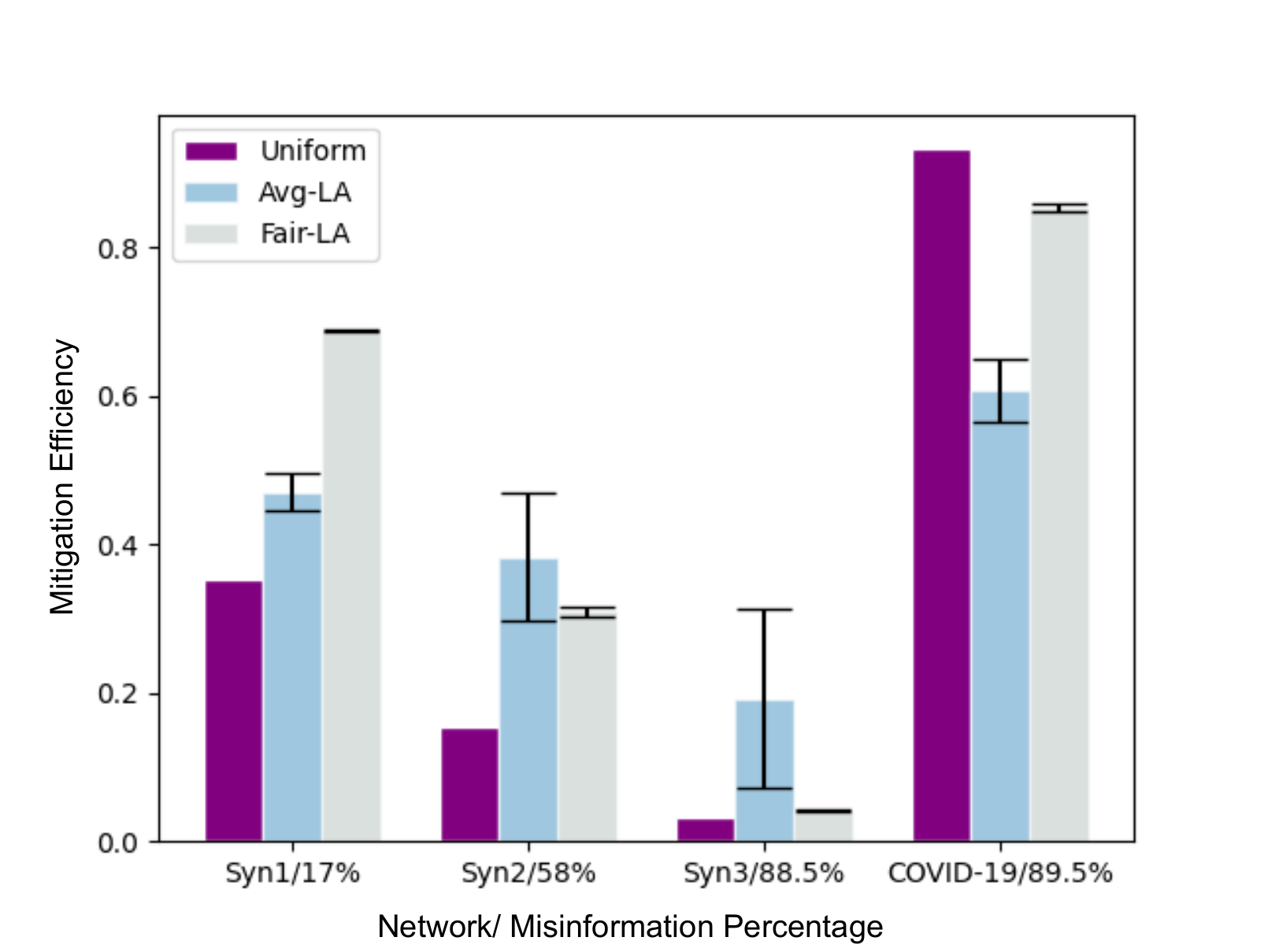}\quad
\includegraphics[height=7cm, width=8.6cm]{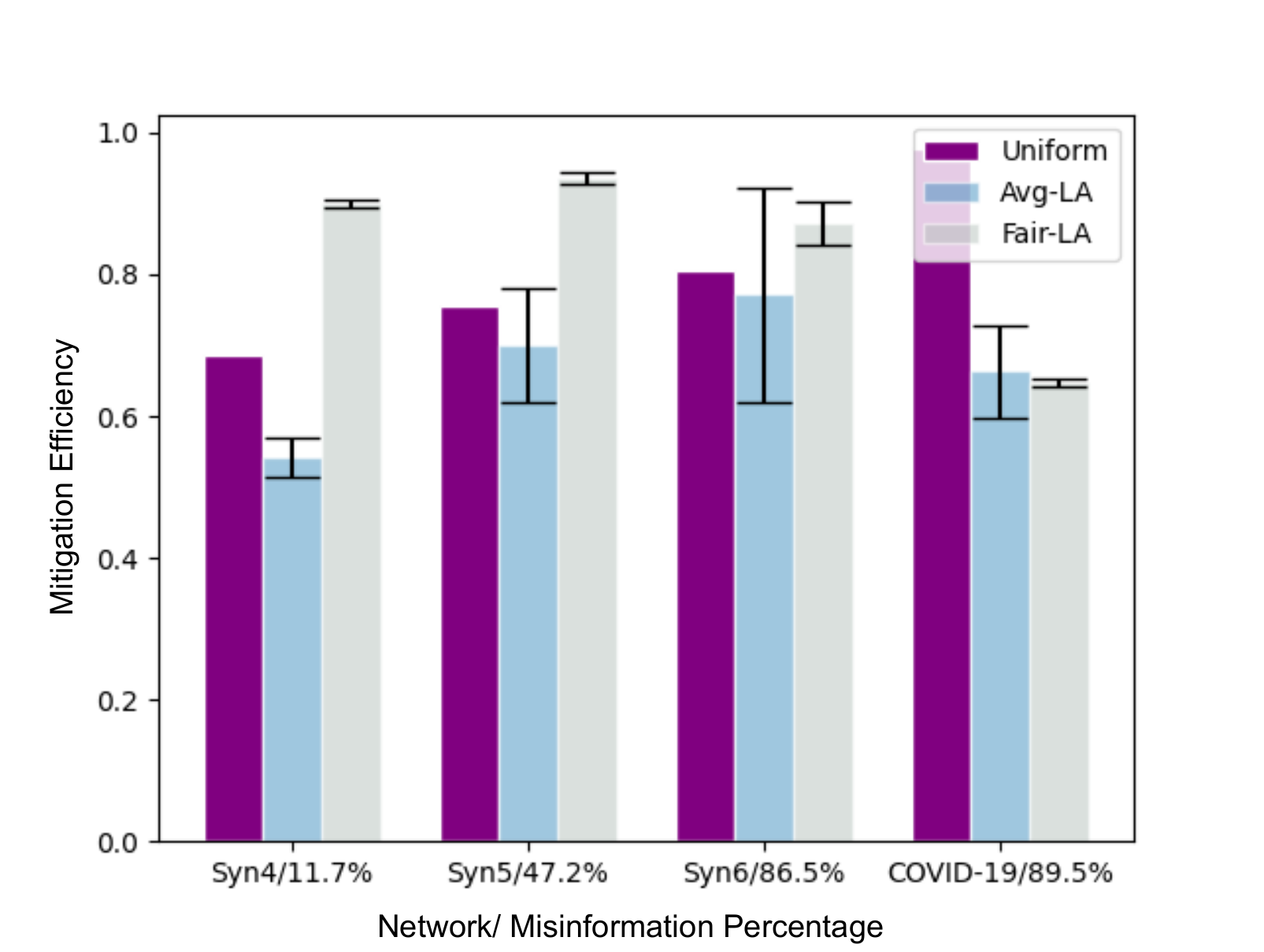}
\caption{Mitigation efficiency on different social network scenarios. Left image: C=0.06, right image: C=0.18.}
\label{fig3}

\end{figure*}

To evaluate for robustness on multiple social networks' scenarios, we introduce a mitigation efficiency metric which is calculated as per the below:
\begin{equation}
\small
    \label{eq14}
    1 - \frac{a}{b}.
\end{equation}
Where $a$ and $b$ are the misinformation percentages after and before mitigation, respectively. According to our synthetic social networks' different setups (See Table \ref{tbl1}), \textbf{Case-1} can be observed in syn1 and syn4, while \textbf{Case-2} can be observed in syn3, and syn6. As concluded from Figure \ref{fig3}, our proposed \textbf{Fair-LA} outperforms both \textbf{AVG-LA} and \textbf{Uniform} methods in most of the scenarios, especially in \textbf{Case-1}. Moreover, when \textbf{Case-2} occurs, \textbf{Fair-LA} still outperforms other methods when the Knapsack capacity $C$ was larger. From our statistical analysis on the COVID-19 network with 200 users, we observed almost a scenario equivalent to \textbf{Case-0}. Therefore, the \textbf{Uniform} method performs better than others. However, we can observe how the efficiency gap is reduced between \textbf{Fair-LA} and \textbf{Uniform} when the Knapsack capacity is more restricted. Eventually, the STD error in the achieved mitigation efficiency percentages for \textbf{Fair-LA} is significantly lower than \textbf{AVG-LA} which also shows how our proposed method is more stable. 

\subsubsection{Fairness Error.} 
\begin{figure*}[h!]
\centering 
\includegraphics[height=7cm, width=8.6cm]{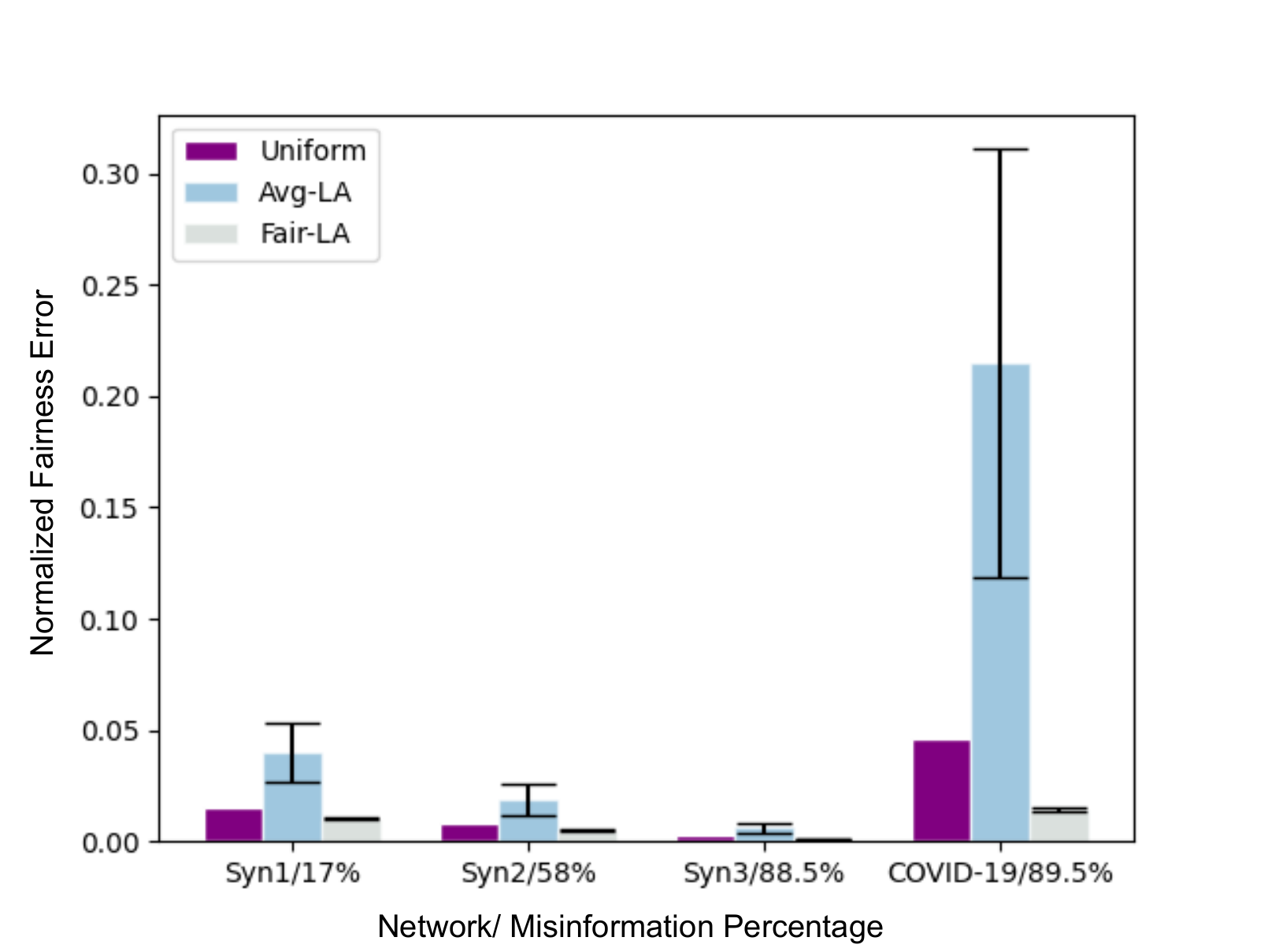}\quad
\includegraphics[height=7cm, width=8.6cm]{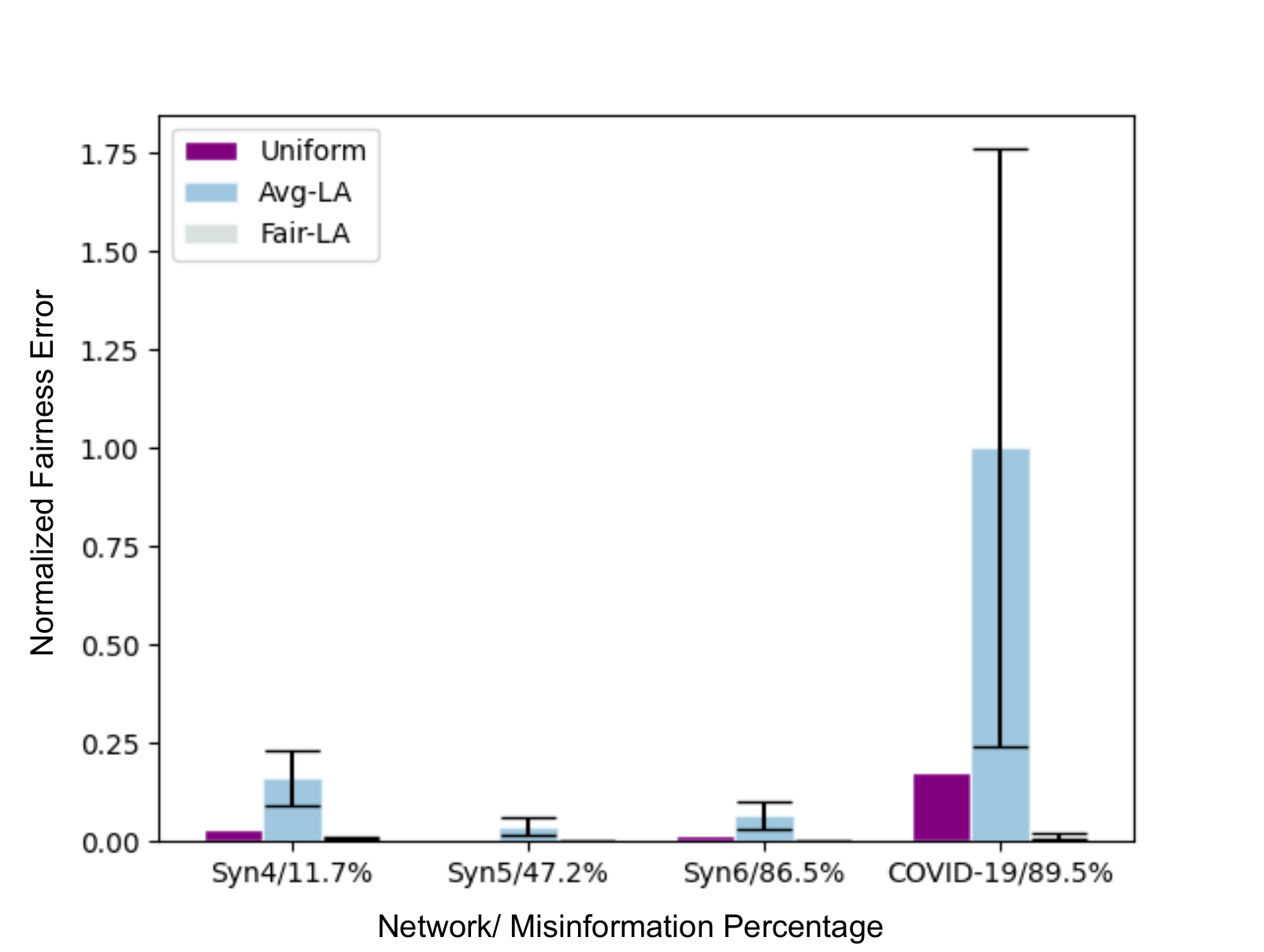}
\caption{Normalized fairness error on different social network scenarios. Left image: C=0.06, right image: C=0.18.}
\label{fig4}
\end{figure*}

Since our proposed loss function (See Equation \ref{eq12}) is considered a general fairness concept, we measure how fair the distribution of incentivization budget among all methods by calculating a normalized total loss. Figure \ref{fig4} shows how our proposed method significantly achieved less fairness error among other methods in all scenarios with stable STD error as well. Consequently, that resulted in not consuming the whole incentivization budget by our method. See Appendix A.4 for more details about how \textbf{Fair-LA} is wisely consuming the Knapsack capacity.

\subsubsection{Learning Bias.}
In the context of our work, a learning bias means unnecessary incentivization values to be assigned based on incomplete evaluation of users' needs due to the non-stationary problem. To reduce such bias, we considered a relatively small learning rate (the automaton state increase/ decrease value) that ensured all users will be visited
almost equal times before consuming the whole budget. Moreover, the fairness error ensured that no user will consume more than
its needs from the budget. Eventually, political polarization would reshape how the learned incentives could actually cause mitigation. Hence, modeling the polarized responses to incentives should be integrated with our resolution in the future work. 

\subsubsection{Desired Mitigation Baseline.} As demonstrated earlier, the idea of misinformation mitigation is to introduce counter information by incentivizing users to propagate it on the network. However, a question remains about to which extend a mitigation should be considered enough. In other words, what if an equal exposure of counter information to misinformation is not enough to maintain authenticity on the network. In such scenario, we propose a balance factor parameter, where the ratios in Equation \ref{eq12} are considered fair only when approaching some balance. For instance, if the desired counter information exposure needed to be twice the amount of misinformation exposure per each user, then, the balance factor is set to $2$ and the fairness of the ratio $R$ is interpreted accordingly. See Appendix A.2.  
\subsubsection{Computation Speed.} Due to the criticality of the misinformation problem, time is an important factor when evaluating misinformation mitigation resolutions. The complete comparison between \textbf{AVG-LA} and \textbf{Fair-LA} regarding their computation speed is given in Appendix A.5.

\subsubsection{Large Scaled Networks.}
Appendix A.6 demonstrates how our method could be scaled on larger networks when sampling techniques are adopted to reduce the optimization space without sacrificing the mitigation efficiency.

\section{Conclusion}
This paper proposed a socially fair approach to misinformation mitigation on social networks. We introduced different synthetic social networks to generate diversity in scenarios where fairness will be critical to how we consume mitigation resources. Unlike other methods, where the fairness perspective was not considered and therefore the social networks which were evaluated were not diverse enough. However, as a limitation in our work, we did not consider the problem of non-responding users in a detailed manner. For instance, some users might be extremely polarized to respond to our incentivization even if their associated HP was responsive. Therefore, we believe that a model for political polarization can be integrated with our proposed method in the future.
\bibliography{main.bib}

\section{Appendix A}
\subsection{A.1 Hawkes Process (HP) Simulation}
To simulate the dynamics of information diffusion on the social network, each user $i$ is associated with two HPs. Therefore, timestamps of both misinformation and normal\footnote{In the domain of misinformation mitigation, some literature refer to the opposite content type of misinformation as normal or true content. In this appendix, we will adopt the term normal content.} content events are considered as inputs for each HP. Then, each HP parameter estimation algorithm \cite{ozaki1979maximum} is responsible for estimating the parameters needed for simulating future events. Hence, the latter estimates the base intensity $\mu$, and the influence matrix $A$ for the kernel function $g$ to calculate the conditional base intensity function as given in Equation \ref{eq1}.

Eventually, the HP thinning algorithm \cite{ogata1981lewis} simulates the network dynamics for the desired time realizations in future. The HP simulation generates future prediction of each user's timestamps for either misinformation or normal content. Across all network's users, we obtain a multivariate Hawkes process (MHP) which describes either the misinformation or the normal content diffusion on the network. The MHP simulation performance is measured according to the below Equation as applied in \cite{abouzeid2021learning} and \cite{farajtabar2017fake}:

\begin{equation}
    \label{a_eq2}
   \mathcal{E}_{t_s + \Delta}:= \frac{1}{N} \sum^{N}_{i=1}|[\lambda^{\mathcal{H}}_{i}(t_s + \Delta) - \lambda^{\mathcal{H}}_{i}(t_s)]-  [\lambda^{\mathcal{R}}_{i}(t_s + \Delta) - \lambda^{\mathcal{R}}_{i}(t_s)]|.
\end{equation}

Where $\mathcal{E}_{t_s + \Delta}$ is the average absolute difference error between actual and predicted timestamps within a certain time realization $t_s$. The actual timestamps (from test data) and the HP predicted timestamps are represented by the intensity functions $\lambda^{\mathcal{R}}_{i}$ and $\lambda^{\mathcal{H}}_{i}$, respectively. We used an exponential decay kernel function $g=A_{i.} e^{-wt}$ as practiced by \cite{abouzeid2021learning} and \cite{farajtabar2017fake}, where $w$ is the decay factor represents the rate for how the influence is reduced over time. We set the intensity decay factors $w=.7$ and $w=1$ for the misinformation and normal content, respectively.

\begin{figure}[t]
\centering
\includegraphics[width=0.55\columnwidth, height=200pt]{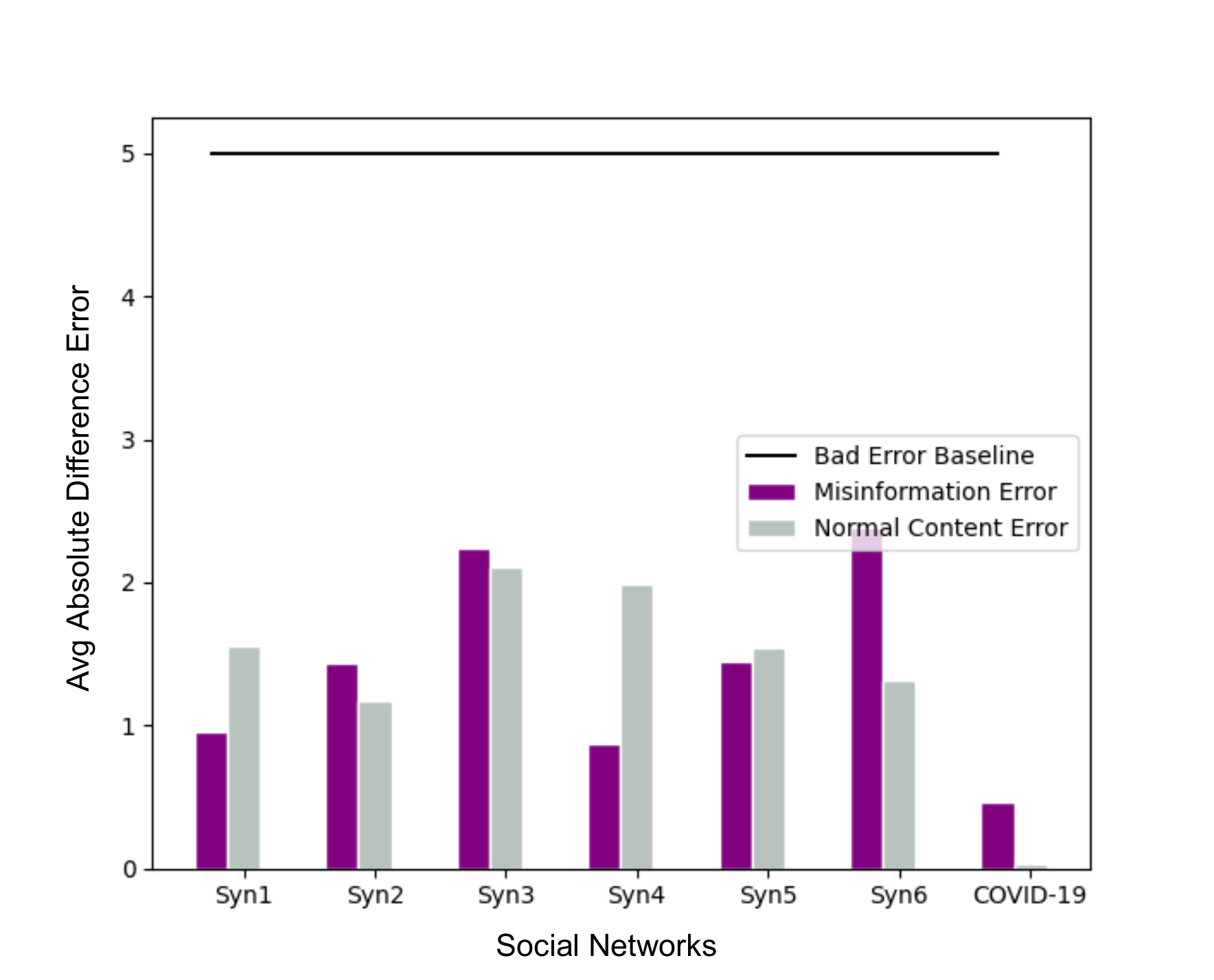} 
\caption{A two-hours realization interval average absolute difference error of all social networks used in the paper experiments.}
\label{a_fig1}
\end{figure}

For any simulated social network, the MHP simulation average absolute difference error is crucial to indicate by how far that simulation is accurate. Consequently, it also indicates how reliable the results from the learned mitigation strategy. Figure \ref{a_fig1} demonstrates the obtained simulation errors on the seven social networks used in our experiments, where time realization intervals are set to two hours, and the task is to predict next two hours events. The error is calculated by splitting the original timestamps to train and test data. In the train data, the HP estimation algorithm is fed with eight hours of users events history on the network. On the other hand, test data contained the next two hours in future, which to be compared with the HP predictions to calculate the average absolute difference error. Eventually, we set an error baseline to $5$, as a baseline for our simulation performance. The error baseline value is inspired by the achieved simulation error in the related work \cite{abouzeid2021learning} and \cite{farajtabar2017fake}. However, we believe our achieved simulation error is significantly small due to the small size of the networks. For the Hawkes parameters estimation and process simulation, we utilize the python package Tick \cite{2017arXiv170703003B} in our python implementation.

\subsection{A.2 User Exposures Ratio}
The weight of either misinformation or normal content on the social network could be viewed as a count. Such counts represent how much each type of content was generated on the network by each user and at a specific time. Moreover, by considering the influence matrix $A$, the influencing users $A_{.i}$ on each user $i$ indicate the amount of exposure (impact) the latter has. Hence, to calculate the impact of network circulated misinformation on user $i$, we sum the number of Hawkes process (HP) generated misinformation events for user $i$ and the number of HP generated misinformation events for its $n$ influencing users derived from $A_{.i}$. For each discrete time realization, the impact is calculated and accumulated with the previous time realizations impact(s). Similarly, we repeat the procedure for the HP generated normal content events. Eventually, when user $i$ is incentivized by the amount $x_i$, the normal content HP base intensity $\mu_i$ is increased by $x_i$, and $R^{x_i}$ is the outcome ratio between the normal content impact (after incentivization) and the default misinformation impact (original HP with no modification). The latter amounts are represented by $T_i(x_i)$ and $F_i$, respectively. Below, we give the complete details of how users impacts ratios are calculated:

\begin{equation}
    \label{a_eq3}
    F^{t_r}_{i}:= \sum^{t_r}_{s=0} \sum^{n}_{j=1} A_{ij} \cdot F^{t_s}_{j},
\end{equation}

\begin{equation}
    \label{a_eq4}
        T^{t_r}_{i}(x_i):= \sum^{t_r}_{s=0} \sum^{n}_{j=1} A_{ij} \cdot T^{t_s}_{j}(x_i),
\end{equation}

\begin{equation}
    \label{a_eq5}
        R^{t_r}_{i}(x_i):= \frac{1 + T^{t_r}_{i}(x_i)}{1 + F^{t_r}_{i}}.
\end{equation}

Where $A_{ij}=1$ if user $i$ is influenced by user $j$, and $A_{ij}=0$ if not, given that user $i$ is considered to be an influencer to itself. $t_s$ is the current time realization index of $r$ discrete realizations and $x_i$ is less than or equal to the maximum allowed incentivization budget $C$. Eventually, we add the value $1$ to both the nominator and denominator of the ratio $R^{t_r}_{i}(x_i)$ to avoid division by zero. Since the definition of a good mitigation is contextual and follows a particular preference, we introduce a balance factor hyper-parameter $b$ which is responsible for determining when to consider the ratio $R$ value as good or not for the mitigation. That is by multiplying the value of the ratio $R$ denominator by $b$. For example, if the impact of normal content was $1$ and the impact of misinformation was also $1$, but we aim to mitigate the misinformation impact with regard to being exposed to normal content by at least $1.5$ more than misinformation, hence the balance factor $b=1.5$. Below is how the impact ratios are calculated, given the balance factor $b$, while we set $b=1.3$ in all experiments done for this paper:

\begin{equation}
\small
    \label{a_eq6}
        R^{t_r}_{i}(x_i):= \frac{1 + T^{t_r}_{i}(x_i)}{(1 + F^{t_r}_{i}) \cdot b} .
\end{equation}

\subsection{A.3 Model Hyper-parameters}
Our proposed misinformation mitigation framework depends on some hyper-parameters for both the information diffusion and the mitigation algorithm. We adopt same information diffusion model parameters as applied in \cite{abouzeid2021learning} and \cite{farajtabar2017fake}. On the other hand, We conduct a grid search to obtain the best parameters values for our proposed mitigation algorithm. Moreover, since the baseline model \cite{abouzeid2021learning} also depends on even more number of hyper-parameters, and was only evaluated on the COVID-19 network, we perform another grid search to obtain its best parameters values on the six proposed synthetic networks for a fair comparison. Table \ref{a_tbl1} shows the final selected values for each model parameter, where our proposed model depends on less hyper-parameters. We refer to our model as \textbf{Fair-LA}, while the other baseline model as \textbf{AVG-LA}. Furthermore, for \textbf{AVG-LA}, we select sample size $|U^-|=200$ which is the whole network, to compare with the maximum performance \textbf{AVG-LA} can achieve. However, in the original \textbf{AVG-LA} experiments, sample size was reduced to $5$ and $25$ as a trade-off between reducing computation time while still achieving the accepted mitigation efficiency.  

 \begin{table}[t]
  \caption{Hyper-parameters final selected values for our model, the baseline model, and the information diffusion model.}
  \centering
  \begin{tabular}{ccccc}
      \hline
    \textbf{Model/Network} & \textbf{Parameter}& \textbf{Value} & \textbf{Description}\\
    \hline
        Fair-LA/all & $b$ & 1.3& balance factor \\
        Fair-LA/all & M&300 &memory depth\\
        AVG-LA/all & M &50 & memory depth\\
        AVG-LA/all & $|U^-|$ &200 &sample size \\
        AVG-LA/Synthetic & $\eta$ &0.0001 & update factor\\
        AVG-LA/COVID-19 & $\eta$ &0.001 & update factor\\
        MHP/all & $w$ &0.7 & mis-decay\\
        MHP/all & $w$ &1 & norm-decay \\
        MHP/all & $\Delta t$ &2 hours & time interval \\

        \hline
  \end{tabular}
  \label{a_tbl1}
\end{table}

\subsection{A.4 Knapsack Budget Consumption}
The \textbf{Fair-LA} is considered a wiser LA when consuming the Knapsack budget. As a consequence, the Knapsack maximum allowed budget $C$ is not always fully consumed since it could happen that no more users need incentivization or their associated HP stopped changing while increasing the base intensity $\mu$. Table \ref{a_tbl2} demonstrates the achieved mitigation efficiency for both \textbf{Fair-LA} and \textbf{AVG-LA}, while Knapsack budget consumption is also given to indicate how with the least amount of incentivization, our proposed \textbf{Fair-LA} is still performing much better in most of the scenarios. 

 \begin{table*}[t]
  \caption{Mitigation efficiency with Knapsack consumption.}
  \centering
  \begin{tabular}{ccccc}
      \hline
    \textbf{Model} & \textbf{Network}& \textbf{Mitigation Efficiency} & \textbf{Efficiency STD Error} & \textbf{c/C}\\
    \hline
        Fair-LA& Syn1 & 0.69 & 0.002 & 0.049\textbf{ /0.06}\\
        AVG--LA& Syn1 & 0.47 &0.025 & 0.059\textbf{ /0.06}\\
        \hline
        Fair-LA & Syn2 &0.31 & 0.007& 0.059\textbf{ /0.06}\\
        AVG-LA & Syn2 &0.38 &0.086 & 0.059\textbf{ /0.06}\\
        \hline
        Fair-LA & Syn3 &0.04 & 0.002& 0.059\textbf{ /0.06}\\
        AVG-LA & Syn3 &0.19 &0.120 & 0.059\textbf{ /0.06}\\
        \hline
        Fair-LA & COVID-19 &0.85 & 0.005& 0.051\textbf{ /0.06}\\
        AVG-LA & COVID-19 &0.61 &0.042 & 0.059\textbf{ /0.06}\\
        \hline
           Fair-LA& Syn4 & 0.90 & 0.006 &0.069\textbf{ /0.18}\\
        AVG--LA& Syn4 & 0.54 &0.028&0.179\textbf{ /0.18}\\
        \hline
        Fair-LA & Syn5 &0.94 & 0.008&0.129\textbf{ /0.18}\\
        AVG-LA & Syn5 &0.70 &0.081 &0.179\textbf{ /0.18}\\
        \hline
        Fair-LA & Syn6 &0.87 & 0.030&0.177\textbf{ /0.18}\\
        AVG-LA & Syn6 &0.77 &0.151 &0.180\textbf{ /0.18}\\
        \hline
        Fair-LA & COVID-19 &0.65 & 0.006&0.089\textbf{ /0.18}\\
        AVG-LA & COVID-19 &0.66 &0.065 &0.179\textbf{ /0.18}\\

        \hline
  \end{tabular}
  \label{a_tbl2}
\end{table*}

\subsection{A.5 Computation Speed}
We run all experiments on a regular CPU workstation with $8$ GB of RAM and Intel Core $i5-8250 @ 1.60 GHz-1.80 GHz$. Table \ref{a_tbl3} compares both \textbf{Fair-LA} and \textbf{AVG-LA} with regard to the computation time of each experiment's single run, averaged over the multiple runs. From the reported statistics in Table \ref{a_tbl3}, we observe how the \textbf{Fair-LA} method high performance comes with a sacrifice to computation speed and convergence time, while we keep observing how the \textbf{Fair-LA} is more stable with regard to the STD error in most of the scenarios. However, in Appendix A.6, we show further enhancements on how a sampling technique improves the \textbf{Fair-LA} computation time.

 \begin{table*}[t]
  \caption{Computation times for both \textbf{Fair-LA} and \textbf{AVG-LA} on all experiments.}
  \centering
  \begin{tabular}{cccc}
      \hline
    \textbf{Model} & \textbf{Network} & \textbf{Time (Hours)}& \textbf{STD Error}\\
    \hline
        Fair-LA& Syn1 & 1.10 & 0.061 \\
        AVG--LA& Syn1 & 0.97 &0.136\\
        \hline
        Fair-LA & Syn2 &1.47 & 0.045\\
        AVG-LA & Syn2 &0.86 &0.132\\
        \hline
        Fair-LA & Syn3 &2.64 & 0.270\\
        AVG-LA & Syn3 &0.76 &0.141 \\
        \hline
        Fair-LA & COVID-19 &3.78 & 0.388\\
        AVG-LA & COVID-19 &1.24 &0.040\\
        \hline
           Fair-LA& Syn4 & 1.09 & 0.062 \\
        AVG--LA& Syn4 & 1.35 &0.421\\
        \hline
        Fair-LA & Syn5 &1.64 & 0.164\\
        AVG-LA & Syn5 &1.24 &0.389 \\
        \hline
        Fair-LA & Syn6 &3.06 & 0.146\\
        AVG-LA & Syn6 &1.13 &0.404 \\
        \hline
        Fair-LA & COVID-19 &4.05 & 0.402\\
        AVG-LA & COVID-19 &1.59 &0.413\\
        \hline
  \end{tabular}
  \label{a_tbl3}
\end{table*}

\subsection{A.6 Sampling}
\begin{figure*}[h!]
\centering
\includegraphics[width=1\columnwidth, height=190pt]{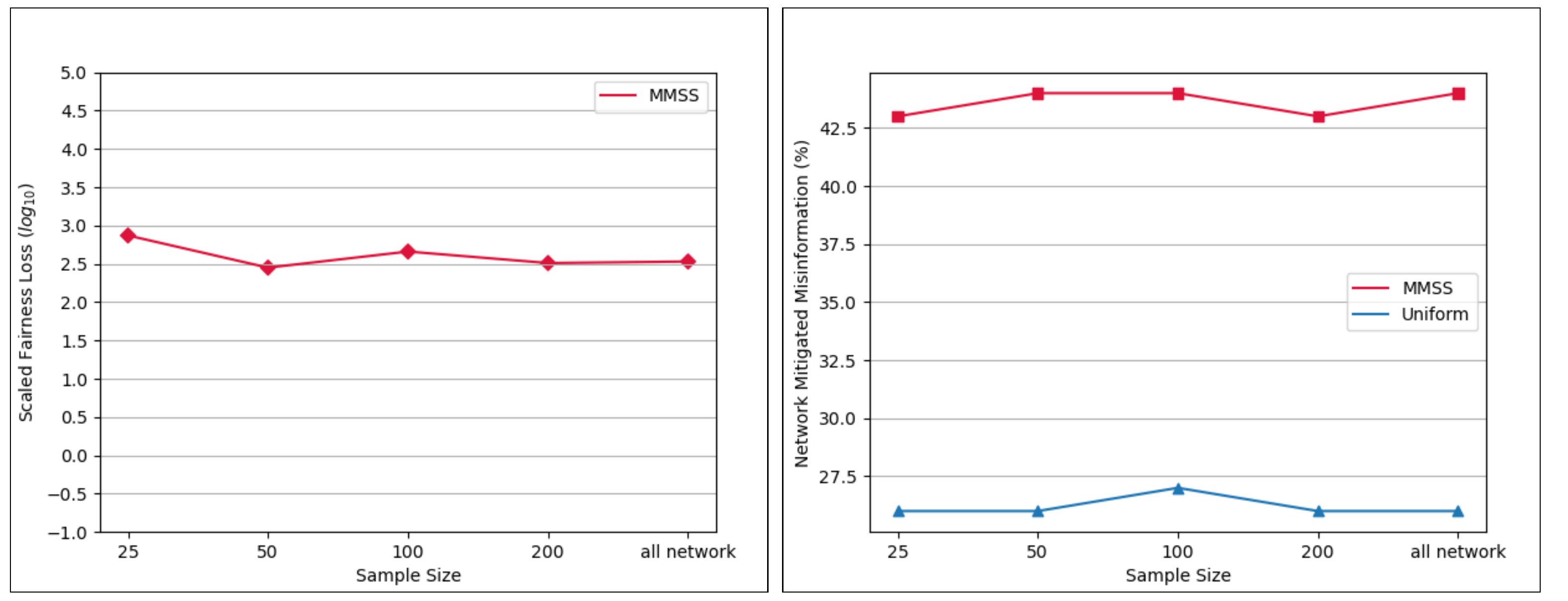} 
\caption{Total fairness loss and mitigation efficiency results on Covid-19 data with sampling.}
\label{fig5}
\end{figure*}

\begin{figure*}[h!]
\centering
\includegraphics[width=1\columnwidth, height=190pt]{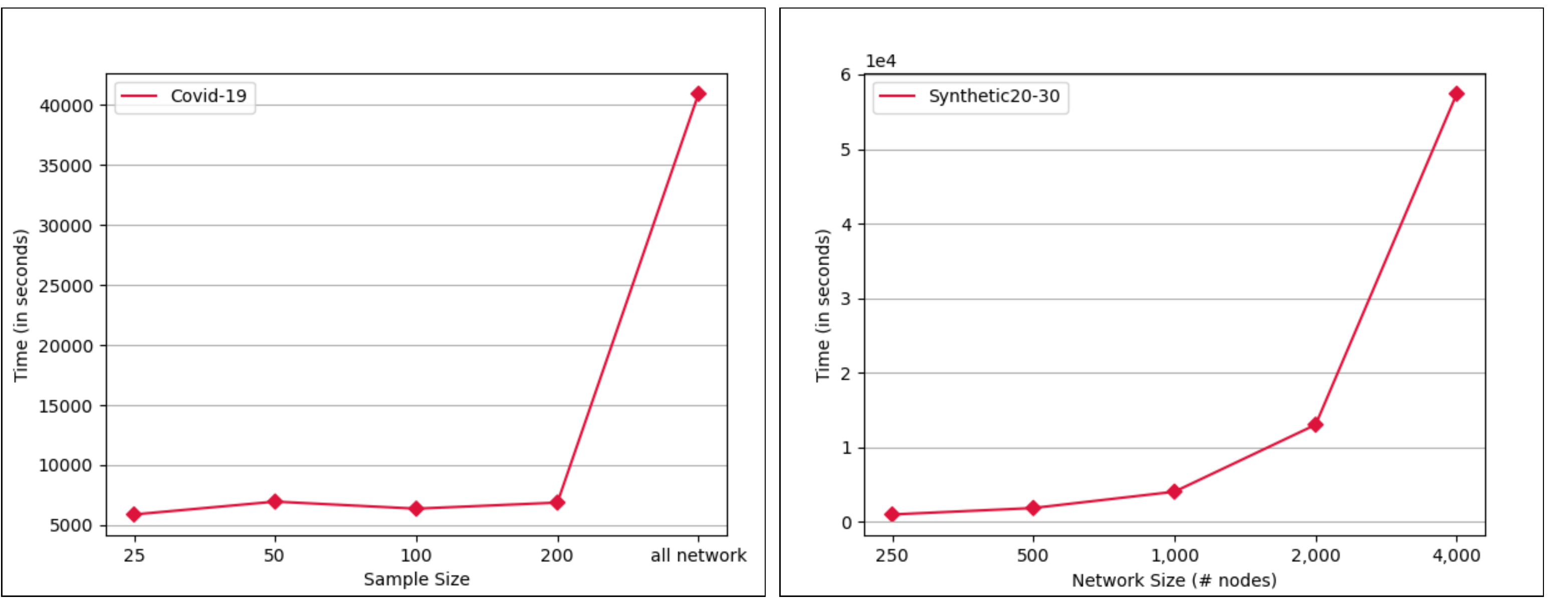} 
\caption{Computation time on different sample and network sizes.}
\label{fig6}
\end{figure*}

Figure \ref{fig5} demonstrates how the \textbf{Fair-LA} mitigates misinformation on a Covid-19 network with $1,164$ users, compared to a poor mitigation by the Uniform method. The sampling technique shows how it did not affect the mitigation efficiency and total fairness loss. The sampling works by evaluating a randomly sampled subset of users for the fairness loss function given in Equation \ref{eq12} instead of all users. We refer to our sampling misinformation mitigation system as \textbf{MMSS-Fair-LA}. For a relatively larger networks' experiments, we use a machine with 64 CPU cores and 128GB of RAM. Figure \ref{fig6} shows computation time and gives an idea of the time complexity of the \textbf{MMSS-Fair-LA} when different sample sizes were used for the whole Covid-19 network. Moreover,  different network sizes were evaluated for the Synthetic networks with misinformation percentages around $20-30$ with fixed sample size of $200$. However, we believe the optimum way to run the \textbf{MMSS-Fair-LA} is on a cluster of machines since the utilized LA network facilitate distributed computing.

\end{document}